\documentclass[aps,prb,groupedaddress]{revtex4}
\usepackage[dvips]{graphicx}

\usepackage{enumerate}

\newcommand{ \Eq  }[1]{Eq.~(\ref{Eq:#1})}
\newcommand{ \Eqs }[2]{Eqs.~(\ref{Eq:#1}) and (\ref{Eq:#2})}

\newcommand{ \Sec     }[1]{Sec.~\ref{Sec:#1}}

\newcommand{ \Figure  }[1]{Figure \ref{Fig:#1}}
\newcommand{ \Fig     }[1]{Fig.~\ref{Fig:#1}}

\newcommand{ \Table }[1]{Table \ref{Table:#1}}

\newcommand{ \Ref  }[1]{Ref.~\onlinecite{#1}}
\newcommand{ \Refs }[2]{Refs.~\onlinecite{#1} and \onlinecite{#2}}

\newcommand{ \etal }{\textit{et al.}}

\newcommand{ \xvec }{\mathbf{x}}
\newcommand{ \rvec }{\mathbf{r}}
\newcommand{ \Rvec }{\mathbf{R}}

\newcommand{ \hydrogen }{(H$_2$)$_{22}$~{}}
\newcommand{ \neon }{Ne$_{13}$~{}}

\newcommand{ \del }{\partial}

\newcommand{ \epsT }{\varepsilon_T}

\newcommand{ \xcent }{\mathbf{x}_\mathrm{c}}

\begin{document}

%\preprint{}

%----------------------------------------------------------------------


\title{Path integral virial estimator based on the scaling of fluctuation coordinates: Application to quantum clusters with fourth-order propagators}

%----------------------------------------------------------------------



%----------------------------------------------------------------------

\author{Takeshi M. Yamamoto}

\email{yamamoto@kuchem.kyoto-u.ac.jp}

\affiliation{Department of Chemistry, Graduate School of Science, Kyoto University, Kyoto 606-8502, Japan}

%----------------------------------------------------------------------


%\date{\today}

%----------------------------------------------------------------------

\begin{abstract}
We first show that a simple scaling of fluctuation coordinates defined in terms of
a given reference point gives the conventional virial estimator
in discretized path integral, where different choices of the reference
point lead to different forms of the estimator
(e.g., centroid virial). 
The merit of this procedure is that it allows
a finite difference evaluation of the virial estimator with respect to
temperature, which totally avoids the need of higher-order potential derivatives.
We apply this procedure to energy and heat capacity calculation
of the \hydrogen and \neon clusters at low temperature using the
fourth-order Takahashi-Imada and Suzuki propagators.
This type of calculation requires up to third-order potential
derivatives if analytical virial estimators are used,
but in practice only first-order derivatives suffice
by virtue of the finite difference scheme above.
From the application to quantum clusters, we find that
the fourth-order propagators do improve upon the primitive
approximation, and that the choice of the reference point
plays a vital role in reducing the variance of the virial estimator.
\end{abstract}

%----------------------------------------------------------------------

%\keywords{}

\maketitle

%\section{}
%\subsection{}
%\subsubsection{}

% If in two-column mode, this environment will change to single-column
% format so that long equations can be displayed. Use
%\begin{widetext}
%\end{widetext}

%----------------------------------------------------------------------

\section{Introduction}

Imaginary time path integral provides a robust way
for studying quantum statistical mechanics of many-particle
systems.\cite{FH65,DMC95}
In the framework of discretized path integral, this
method maps a quantum system into multiple copies of virtual classical systems
(called ``beads'') connected via harmonic springs. This isomorphism
allows one to calculate structural and thermodynamic properties using
conventional Monte Carlo or molecular dynamics methods. In practice,
however, such calculation often becomes much more demanding
than the classical counterpart, and thus a number of efficient
techniques have been developed, e.g.,
collective sampling of multiple beads,\cite{DMC95,PC84,SKC85,TBMK93}
statistical estimators with low variance,\cite{B79,HBB82,PR84,CB89,JS97,EDCF99,GF02a,GF02b,NFD00,PSDF03_Ne,SS05}
and accurate approximations to the exact short-time
propagator (or high-temperature density matrix).\cite{PC84,S95,C97,TI84,LB87,MM89,MMB00,JJV01,CC01,WKEC98,BLL03,KM02,CMMB03,BSBC04,P04a,C04,C05,CB92,SL95,BBB05}

Our interest in this paper is in the latter two issues, namely the use of better
statistical estimators and approximate propagators.
Regarding the estimator, the most conventional path-integral
estimators for internal energy are thermodynamic \cite{B79} and
virial\cite{HBB82} estimators.
The former is obtained via direct temperature differentiation of
the partition function, while the latter is obtained by eliminating
ill-behaved terms in the former through integration by parts.
The virial estimator has an advantage that
its variance is only weakly dependent on the number of beads, $P$,
in contrast to the thermodynamic estimator whose variance grows linearly with $P$.
We emphasize, however, that this reduction in the variance
is achieved at the expense of using
first-order potential derivatives that are absent
in the thermodynamic estimator. Although the
first-order derivatives are often not a major computational problem,
things become worse when one constructs a similar \textit{double} virial
estimator for heat capacity because
it requires second-order potential
derivatives. Despite this difficulty,
the double virial estimator was
used in a heat capacity calculation of water because other estimators
exhibited too large statistical errors and could not be converged within simulation
time.\cite{SS05}
To remedy this problem, Glaesemann and Fried\cite{GF02a,GF02b}
proposed a free-particle projection technique to reduce the variance of
the thermodynamic estimator without using potential derivatives, and applied it
to Ar$_6$ clusters with considerable success at higher temperature.
Predescu \etal\cite{PSDF03_Ne} adopted a different strategy in their random series
path integral (generalized form of the Fourier path integral), where
they first scaled the amplitude of the Brownian
bridge and then differentiated the scaled partition function via finite difference
in order to obtain a viriallike estimator having no potential derivatives.
With this method they calculated
the quantum heat capacity of the \neon cluster at 4--14 K
with unprecedented accuracy.\cite{PSDF03_Ne}

Another issue that impacts the efficiency of path integral is
the accuracy of approximate propagators.
There exist a number of such approximations that aim at faster
convergence to the $P \rightarrow \infty$ limit
than the standard primitive approximation.
In particular, the pair-product approximation\cite{DMC95,PC84}
and the higher-order composite factorizations\cite{TI84,S95,C97,JJV01}
have proven to be successful in
condensed-phase applications (see \Ref{KM02} for their useful comparisons). The
pair-product approximation replaces the exact high-temperature density matrix
by the product of effective pairwise ones, and it has been shown to
drastically reduce the number of beads for
monoatomic fluids.\cite{SL95,CGC98}
The fast convergence of this approximation was also exploited
in semiclassical dynamical calculation of
normal and superfluid helium.\cite{NM03}
While powerful for monoatomic fluids,
the pair-product approximation becomes cumbersome when applied to
molecular fluids because of the increased complexity of pair action.
In this regard the higher-order propagators are appealing in that molecular
fluids can be treated straightforwardly.
In practice, however,
the application of such propagators to molecular fluids
is very scant compared to the primitive approximation.
One reason may be that the higher-order propagators involve the
first-order potential derivatives, and
the corresponding virial estimator for energy and heat capacity requires
second- and third-order potential derivatives, respectively,
resulting in a significant computational overhead.
(Incidentally, Jang \etal\cite{JJV01}
showed that for the Suzuki propagator the required order of potential derivatives
can be reduced by using the virial theorem in operator form.)

In this paper we present a method for evaluating the virial and double
virial estimators in discretized path integral
without using higher-order potential derivatives. This
method is based on the coordinate scaling idea of Janke and Sauer\cite{JS97} and
the finite difference method of Predescu \etal.\cite{PSDF03_Ne}
Specifically, we first show
that a simple scaling of fluctuation coordinates defined in terms of a given
reference point gives the conventional virial estimator, where different
choices of the reference point lead to different forms of the estimator
(e.g., centroid virial). This procedure reverts to the original
coordinate scaling by Janke \etal~when the reference point is
set to the coordinate origin.
We then take the temperature derivative of the scaled partition function
by finite difference in order to avoid potential derivatives.
We illustrate the above method by calculating energy and heat
capacity of the \hydrogen and \neon clusters
at low temperature using the fourth-order
composite propagators. This calculation requires up to third-order potential
derivatives if analytical virial estimators are used, but in practice
only up to first-order derivatives suffice by virtue of the finite difference
scheme above. From the results of the application,
we find that the fourth-order propagators
do improve upon the primitive approximation, and
that the choice of the reference point has a crucial role
in reducing the variance of the virial estimator.

The remainder of this paper is as follows: In \Sec{method} we describe
the coordinate scaling and finite difference procedures mentioned above.
In \Sec{application} we apply the present method to
the \hydrogen cluster at 6 K and \neon cluster at 4--14 K and
calculate their total energy, heat capacity, and distance distribution
functions. Systematic comparisons
are made among different types of propagators and estimators.
In \Sec{conclusion} we conclude.

%----------------------------------------------------------------------

\section{Path integral estimators for energy and heat capacity\label{Sec:method}}

\subsection{Conventional estimators}

We first summarize the conventional thermodynamic\cite{B79}
and virial\cite{DMC95,HBB82,PR84} estimators for
subsequent discussion. We suppose an $f$-dimensional system having the
Hamiltonian $H = T + V = \sum_{i=1}^{f} p_i^2 / 2 m + V ( \xvec )$ with
$\xvec = ( x_1, \ldots, x_f )$.
Using the primitive approximation to the canonical density operator,
\begin{equation}
  e^{- \epsilon \hat H} = e^{- \epsilon \hat V / 2} e^{- \epsilon \hat T}
  e^{- \epsilon \hat V / 2} + O ( \epsilon^3 ),
\end{equation}
the partition function at inverse temperature $\beta = 1 / k_B T$ can be
written as
\begin{equation}
\label{Eq:Z_PA}
  Z ( \beta ) = \mathrm{tr} ( e^{- \beta \hat H} ) = \int d \xvec_1
 \cdots \int d \xvec_P
  \rho ( \xvec_1, \ldots, \xvec_P ; \beta ) + O ( 1 / P^2 )
\end{equation}
with
\begin{equation}
\label{Eq:rho_PA}
  \rho ( \xvec_1, \ldots, \xvec_P ; \beta ) = \left( \frac{m P}{2 \pi \hbar^2
  \beta} \right)^{ P f / 2 } \exp \left\{ - \frac{m P}{2 \hbar^2 \beta} \sum^P_{s
  = 1} ( \xvec_s - \xvec_{s - 1} )^2 - \frac{\beta}{P} \sum^P_{s = 1} V ( \xvec_s )
  \right\},
\end{equation}
where $\xvec_s$ is the system coordinate in the $s$-th
time slice (or ''beads'') with $\xvec_0 = \xvec_P$. The thermodynamic estimator is
obtained by direct temperature differentiation of \Eq{Z_PA}:
\begin{equation}
  E ( \beta ) = - \frac{1}{Z ( \beta )} \frac{\partial Z ( \beta )}{\partial
  \beta} \simeq \langle \varepsilon_T \rangle
\end{equation}
with
\begin{equation}
\label{Eq:eps_T}
  \varepsilon_T = \frac{ P f }{2 \beta} - \frac{m P}{2 \hbar^2 \beta^2} \sum^P_{s
  = 1} ( \xvec_s - \xvec_{s - 1} )^2 + \frac{1}{P} \sum^P_{s = 1} V ( \xvec_s ),
\end{equation}
where $\langle \cdots \rangle$ denotes an ensemble average over
the sampling function $\rho ( \xvec_1, \ldots, \xvec_P ; \beta )$. The drawback of
this estimator is that its variance grows with $P$ due to
cancellation of the first two terms in the right-hand side of \Eq{eps_T}. This
difficulty can be avoided by using the relation,
\begin{equation}
\label{Eq:int_by_parts}
  \int d \xvec_1 \cdots \int d \xvec_P  \left[ \sum^P_{s = 1} ( \xvec_s - \xvec^* )
  \cdot \frac{\partial}{\partial \xvec_s} \right]
  \rho ( \xvec_1, \ldots, \xvec_P ; \beta )
 = - ( P - g ) f \int d \xvec_1 \cdots \int d \xvec_P
  \rho ( \xvec_1, \ldots, \xvec_P ; \beta ),
\end{equation}
which arises from integration by parts. In \Eq{int_by_parts}, $\xvec^*$
 is a given
``reference'' point and $g$ is a constant that depends on the definition of
$\xvec^*$. In this paper we consider three choices of $\xvec^*$, namely $\xvec^*
= 0$, $\xvec_P$, and $\xcent$, where $\xcent$ is the centroid of the
imaginary-time path given by
\begin{equation}
  \xcent = \frac{1}{P} \sum_{s = 1}^P \xvec_s .
\end{equation}
With these choices the value of $g$ becomes\cite{int_by_parts}
\begin{equation}
\label{Eq:g}
  g = \left\{\begin{array}{l}
    0, \quad \xvec^* = 0,\\
    1, \quad \xvec^* = \xvec_P \; \mathrm{and} \; \xcent.
  \end{array}\right.
\end{equation}
Because the kinetic action in \Eq{rho_PA}
is doubled by the ``virial operator'' in the square bracket in 
\Eq{int_by_parts},\cite{Euler_theorem}
the following path integral virial theorem holds:
\begin{equation}
  \left\langle \frac{m P}{2 \hbar^2 \beta^2} \sum^P_{s = 1}
 ( \xvec_s - \xvec_{s - 1} )^2
 + \frac{1}{2 P} \sum^P_{s = 1} ( \xvec_s - \xvec^* )
 \cdot \frac{ \partial V ( \xvec_s ) }{ \partial \xvec_s }
\right\rangle
  = \frac{ (P - g) f }{2 \beta} .
\end{equation}
Eliminating the first two terms in \Eq{eps_T} through the above relation,
we have the following virial estimator for energy:
\begin{equation}
\label{Eq:eps_V}
  \varepsilon_V = \frac{ f g }{2 \beta} + \frac{1}{P} \sum^P_{s = 1} \left[
  \frac{1}{2} ( \xvec_s - \xvec^* ) \cdot
 \frac{ \partial V ( \xvec_s ) }{ \partial \xvec_s }
 + V ( \xvec_s ) \right] .
\end{equation}
For convenience we will refer to the above estimator with $\xvec^* = 0$,
$\xvec_P$, and $\xcent$ as the origin-, bead-, and centroid-reference virial
estimators, respectively. We note that the origin-reference virial
estimator gives an incorrect result for unbounded systems\cite{DMC95} (e.g.,
$\varepsilon_V$ vanishes for a free particle) although the bead- and
centroid-reference virial estimators remain valid. The reason
is that in the former the integral of $\rho ( \xvec_1, \ldots, \xvec_P ; \beta )$ over the whole
coordinate space is divergent, which invalidates \Eq{int_by_parts},
while in the latter
the integration by parts can be performed in one less dimensions with some
coordinate fixed (e.g., $\xvec_P$ in the bead-reference virial). Despite this
deficiency, the origin-reference virial estimator can be applied to
quantum clusters if the contribution of the center of mass is properly
taken into account.\cite{EDCF99,NFD00}

Heat capacity estimators can be obtained in a similar manner and are
fully described in \Ref{GF02b}.
The resulting double thermodynamic estimator contains no
potential derivatives but its variance grows rapidly as $P^2$. The double
virial estimator has a favorable variance weakly
dependent on $P$ but it requires second-order potential derivatives,
resulting in an increased computational effort.\cite{SS05}

%----------------------------------------------------------------------

\subsection{Virial estimator via coordinate scaling\label{Sec:fd}}

The virial estimator in \Eq{eps_V} can also be obtained by a scaling of fluctuation
coordinates as mentioned in the Introduction. This is achieved by first
considering the partition function at a different temperature $\beta'$:
\begin{equation}
\label{Eq:Z_diff_beta}
  Z ( \beta' ) = \int d \xvec'_1 \cdots \int d \xvec'_P \rho ( \xvec'_1, \ldots,
  \xvec'_P ; \beta' ),
\end{equation}
where $\rho$ is the density function in \Eq{rho_PA}. To eliminate ill-behaved
terms in the thermodynamic estimator, we introduce a new set of variables $(
\xvec_1, \ldots, \xvec_P )$ as
\begin{equation}
\label{Eq:scaling}
  \xvec'_s = \xvec^* + \sqrt{\frac{\beta'}{\beta}} ( \xvec_s - \xvec^* ),
\end{equation}
for $s = 1, \ldots, P$. $\xvec^*$ in \Eq{scaling} is a reference point that has the
same meaning as in the preceding section, i.e., $\xvec^* = 0$, $\xvec_P$, or
$\xcent$. The Jacobian of this transformation is
\begin{equation}
  d \xvec'_1 \cdots d \xvec'_P = 
\left( \frac{\beta'}{\beta} \right)^{ ( P - g ) f / 2 } d \xvec_1 \cdots d \xvec_P,
\end{equation}
where $g$ is given by \Eq{g}. Since the transformation in \Eq{scaling} suggests
\begin{equation}
  \frac{1}{\beta'} \sum_{s = 1}^P ( \xvec'_s - \xvec'_{s - 1} )^2 = \frac{1}{\beta}
  \sum^P_{s = 1} ( \xvec_s - \xvec_{s - 1} )^2,
\end{equation}
the partition function in \Eq{Z_diff_beta} may be written as
\begin{equation}
  Z ( \beta' ) = \int d \xvec_1 \cdots \int d \xvec_P
 \rho ( \xvec_1, \ldots, \xvec_P ; \beta ) R ( \beta' )
\end{equation}
with
\begin{equation}
\label{Eq:R}
  R ( \beta' ) = \left( \frac{\beta}{\beta'} \right)^{ f g / 2 } \exp \left\{ -
  \frac{1}{P} \sum_{s = 1}^P [ \beta' V ( \xvec'_s ) - \beta V ( \xvec_s ) ]
 \right\}.
\end{equation}
Using the above equation the internal energy is obtained as follows,
\begin{equation}
  E ( \beta )
  =
  -
  \left.
  \left\langle
     \frac{ \del R( \beta' ) }{ \del \beta' }
  \right\rangle
  \right|_{ \beta' = \beta }
  = 
  \left\langle \varepsilon_V \right\rangle
\end{equation}
with
\begin{equation}
  \label{Eq:eps_V2}
  \varepsilon_V
  =
  \left.
     \frac{ f g }{ 2 \beta }
     + \frac{ 1 }{ P }
     \sum_{ s=1 }^{ P }
     \frac{ \partial }{ \partial \beta' }
     [ \beta' V( \xvec'_s(\beta') ) ]
  \right|_{ \beta' = \beta }
  ,
\end{equation}
where we have explicitly denoted the $\beta'$-dependence of $\xvec'_s$.
$\varepsilon_V$ in \Eq{eps_V2} can be shown identical to
that in \Eq{eps_V} by taking the $\beta'$-derivative analytically.
Instead, we may take the $\beta'$-derivative via finite difference
in order to avoid potential derivatives:\cite{PSDF03_Ne}
\begin{equation}
  \label{Eq:eps_V2_fd}
  \varepsilon_V
  \simeq
  \frac{ f g }{ 2 \beta }
  + \frac{ 1 }{ 2 P \delta\beta } \sum_{ s=1 }^{ P }
  \left[
     ( \beta + \delta\beta ) V( \xvec'_s ( \beta + \delta\beta ) )
   - ( \beta - \delta\beta ) V( \xvec'_s ( \beta - \delta\beta ) )
  \right]
  .
\end{equation}
Similarly, the constant volume heat capacity,
\begin{equation}
  C_V ( \beta )
  =
  \frac{d E(T)}{d T}
  =
  k_B \beta^2 \left\{
  \frac{1}{Z ( \beta )} \frac{\partial^2 Z ( \beta )}{\partial
  \beta^2} - \left[ \frac{1}{Z ( \beta )} \frac{\partial Z (
  \beta )}{\partial \beta} \right]^2 \right\},
\end{equation}
can be obtained using the following expression,
\begin{equation}
  \label{Eq:C_fd}
  C_V ( \beta )
  =
  k_B \beta^2
  \left\{
    \langle \varepsilon_V^2 \rangle
    -
    \langle \varepsilon_V \rangle^2
    - \langle \varepsilon'_V \rangle
  \right\}
\end{equation}
with
\begin{eqnarray}
  \label{Eq:eps_V2_deriv}
  \varepsilon'_V
  & = &
  \left.
     - \frac{ f g }{ 2 \beta^2 }
     + \frac{ 1 }{ P }
     \sum_{ s=1 }^{ P }
     \frac{ \partial^2 }{ \partial \beta'^2 }
     [ \beta' V( \xvec'_s (\beta') ) ]
  \right|_{ \beta' = \beta }
  ,
\end{eqnarray}
where we may use finite difference to evaluate
the second derivative with respect to $\beta'$.
Although there are other schemes for performing finite difference, e.g.,
\begin{equation}
  \label{Eq:fd_R}
  E( \beta )
  \simeq
  -
  \left\langle
     \frac
     { R( \beta + \delta\beta ) - R( \beta - \delta\beta ) }
     { 2 \delta \beta }
  \right\rangle
  ,
\end{equation}
it has a narrower range of acceptable values of
$\delta \beta$ than \Eq{eps_V2_fd}
due to the exponential behavior of $R(\beta')$.
Therefore we will use only \Eqs{eps_V2_fd}{eps_V2_deriv} in the following sections.

We emphasize that the finite difference scheme above is
qualitatively different, e.g., from that performed for internal energy
with respect to temperature,
\begin{equation}
  \label{Eq:C_V_bad_fd}
  C_V ( \beta ) \simeq \frac{E ( T + \delta T ) - E ( T - \delta T )}{2 \delta
  T}
  .
\end{equation}
where $E(T)$ contains statistical error and
$\delta T$ must be taken sufficiently large so that
$|E ( T + \delta T ) - E ( T - \delta T ) |$ is much
larger than the statistical error in $E ( T \pm \delta T )$. On the other hand,
the finite difference in \Eq{eps_V2_fd} is performed for
a statistical error-free quantity, $\beta' V( \xvec'_s ( \beta' ) )$,
and thus $\delta \beta$ can be taken as small as machine precision allows.
In practice, however, acceptable values of $\delta \beta$
may depend on the stiffness of the potential as well as
thermodynamic conditions under study (e.g., particle density),
so one needs to check the convergence
by repeating a very short simulation with different values of $\delta\beta$.
Our typical choice of $\delta\beta$ is $10^{-4}\beta$
(see \Sec{application}).

%----------------------------------------------------------------------

\subsection{Using fourth-order composite propagators\label{Sec:FOA}}

An appealing feature of the finite-difference scheme in \Sec{fd} is that it does
not require potential derivatives higher than those existing in the
discretized action. This means that
energy and heat capacity can be calculated with no potential derivatives
when the primitive approximation is used,
and only up to first-order derivatives are needed when the
fourth-order composite propagators are used. The generalized
Suzuki\cite{S95,C97,JJV01} and Takahashi-Imada\cite{TI84} approximations
fall into the latter category.
The Suzuki approximation factorizes the exact short-time propagator as
\begin{equation}
\label{Eq:Suzuki}
  e^{- 2 \epsilon H} = e^{- \epsilon \tilde{V}_e / 3}
  e^{- \epsilon T} e^{- 4 \epsilon \tilde{V}_m / 3}
  e^{- \epsilon T} e^{- \epsilon \tilde{V}_e / 3}
  + O ( \epsilon^5 ),
\end{equation}
where $\tilde{V}_m$ and $\tilde{V}_e$ are effective potentials that involve
first-order potential derivatives (see \Refs{C97}{JJV01} for details). With this
factorization the approximate partition function becomes
\begin{equation}
  \label{Eq:Z_FOA}
  Z ( \beta ) = \int d \xvec_1 \cdots \int d \xvec_P \rho^{( 4 )} ( \xvec_1,
  \ldots, \xvec_P ; \beta ) + O ( 1 / P^4 )
\end{equation}
with
\begin{equation}
\label{Eq:rho_FOA}
  \rho^{( 4 )} ( \xvec_1, \ldots, \xvec_P ; \beta ) = \left( \frac{m P}{2 \pi
  \hbar^2 \beta} \right)^{ P f / 2 } \exp \left\{ - \frac{m P}{2 \hbar^2 \beta}
  \sum^P_{s = 1} ( \xvec_s - \xvec_{s - 1} )^2 - \frac{\beta}{P} \sum^P_{s = 1} w_s
  \tilde{V}_s ( \xvec_s ; \beta ) \right\},
\end{equation}
where $\tilde{V}_s$ is a time-slice dependent effective potential defined by
\begin{equation}
  \label{Eq:V_eff}
  \tilde{V}_s ( \xvec ; \beta ) = V ( \xvec ) + d_s ( \beta / P )^2 C ( \xvec )
\end{equation}
with
\begin{equation}
\label{Eq:commutator}
  C ( \xvec ) = [ V, [ T, V ] ] = \frac{\hbar^2}{m}
 \left| \frac{ \partial V( \xvec ) }{ \partial \xvec } \right|^2,
\end{equation}
while $w_s$ and $d_s$ are a set of coefficients given by
\begin{equation}
  w_s = \left\{\begin{array}{ll}
    2 / 3, & \; s = \mathrm{even},\\
    4 / 3, & \; s= \mathrm{odd},
  \end{array}\right.
\end{equation}
and
\begin{equation}
  d_s = \left\{\begin{array}{ll}
    \alpha / 6,          & \; s = \mathrm{even},\\
    ( 1 - \alpha ) / 12, & \; s = \mathrm{odd},
  \end{array}\right.
\end{equation}
where $\alpha$ is an arbitrary parameter within [0,1]. The partition function
for the Takahashi-Imada approximation\cite{TI84} can also be expressed
in the form (\ref{Eq:rho_FOA})
with $w_s = 1$ and $d_s = 1 / 24$, although this approximation is not based on
a genuine factorization such as \Eq{Suzuki}.
The fourth-order approximation in \Eq{Z_FOA} differs
from the primitive, second-order one only in that the bare potential is replaced by
the slice-dependent effective potential in \Eq{V_eff}, and that the weight factors
$\{w_s\}$ are introduced in a way similar to
Simpson's quadrature rule. It is thus straightforward to
apply the procedure in \Sec{fd} to obtain finite-difference virial estimators
having no higher-order potential derivatives. Specifically,
the statistical average is now taken over $\rho^{(4)}$ in \Eq{rho_FOA}, and
we modify $R ( \beta' )$ in \Eq{R} as
\begin{equation}
  R ( \beta' )
  =
  \left( \frac{\beta}{\beta'} \right)^{ f g / 2 } \exp \left\{ -
  \frac{1}{P} \sum_{s = 1}^P w_s [ \beta' \tilde{V}_s ( \xvec'_s ; \beta' ) -
  \beta \tilde{V}_s ( \xvec_s ; \beta ) ] \right\}
  ,
\end{equation}
and the virial estimators in \Eqs{eps_V2}{eps_V2_deriv} as follows:
\begin{equation}
  \varepsilon_V
  =
  \left.
     \frac{ f g }{ 2 \beta }
     +
     \frac{ 1 }{ P } 
     \sum_{ s=1 }^{ P } w_s
     \frac{ \partial }{ \partial \beta' }
     [ \beta' \tilde{V}_s ( \xvec'_s ; \beta' ) ]
  \right|_{ \beta' = \beta }
  ,
\end{equation}
\begin{equation}
  \varepsilon'_V
  =
  \left.
     - \frac{ f g }{ 2 \beta^2 }
     + \frac{ 1 }{ P } 
     \sum_{ s=1 }^{ P } w_s
     \frac{ \partial^2 }{ \partial \beta'^2 }
     [ \beta' \tilde{V}_s ( \xvec'_s ; \beta' ) ]
  \right|_{ \beta' = \beta }
  .
\end{equation}

%----------------------------------------------------------------------

\section{Application to quantum clusters\label{Sec:application}}

\subsection{(H$_{2}$)$_{22}$ cluster at 6 K\label{Sec:hydrogen}}

We illustrate the above procedure by first calculating the energy and heat
capacity of the \hydrogen cluster at 6 K. The physical model
is identical to that used in the previous studies.\cite{CGC98,DF99,PSDF03_H2}
Briefly, the system potential consists of Lennard-Jones (LJ) pair interactions with
$\epsilon_\mathrm{LJ} = 34.2$ K and $\sigma_\mathrm{LJ} = 2.96$ \AA, where the
hydrogen molecules are treated as distinguishable spherical particles with
their mass being 2 amu. Since a cluster in vacuum at any
positive temperature is metastable with respect to evaporation, a confining
potential of the form,
\begin{equation}
\label{Eq:V_c}
  V_c
  =
  \epsilon_\mathrm{LJ}
  \sum^{ N }_{i = 1}
  \left(
     \frac{ | \rvec_i - \Rvec | }
          { R_c }
  \right)^{20},
\end{equation}
is added to the sum of LJ potentials to prevent any molecules from permanently
leaving the cluster. In \Eq{V_c}, $\mathbf{r}_i$ is the position of particle
$i$, $\mathbf{R}$ is the center of mass of the cluster, $N$ is the number
of particles, and $R_c$ is the
confining radius chosen as $4 \sigma_\mathrm{LJ}$.

Statistical sampling of imaginary-time paths was performed with Monte Carlo
(MC) methods, where one cycle is defined such that each particle is moved once
on average by the staging algorithm.\cite{PC84,SKC85,TBMK93}
The implementation is the same as
described in \Ref{TBMK93}. The staging length $j$
(the number of beads that are collectively moved)
was determined by adjusting the acceptance ratio to $50$ \%,
which resulted in $j \approx P / 4$ regardless of the value of $P$.
In addition to the staging move, we applied the whole-chain move\cite{DMC95}
every two cycles to accelerate the statistical convergence.
A single run consisted of $5\times10^5$ cycles for equilibration followed by
$4\times10^6$ cycles for data accumulation, which took several days
for $P = 160$ using a Pentium4 3.8 GHz PC.

\begin{table}
\caption{\label{Table:E}
Internal energy (K/molecule) calculated using the centroid
virial estimator based on the finite difference scheme in \Sec{fd}. PA, TIA, and
SA denote the primitive, Takahashi-Imada, and Suzuki approximations,
respectively. $\alpha$ is an arbitrary parameter within [0,1]
involved in the Suzuki approximation.
The figure in parentheses is one standard deviation on the last digit.
}
\begin{ruledtabular}
\begin{tabular}{cccccc}
%----------------------------------------------------------------------
P & PA & TIA & SA($\alpha=0$) & SA ($\alpha$=1/2) & SA($\alpha$=1) \\
%----------------------------------------------------------------------
\hline
   20 & $-$27.52(1) & $-$21.80(1) & $-$23.24(1) & $-$20.94(1) & $-$21.66(1) \\
   40 & $-$21.54(1) & $-$18.80(1) & $-$19.41(1) & $-$18.34(1) & $-$18.38(1) \\
   60 & $-$19.78(1) & $-$18.13(1) & $-$18.41(1) & $-$17.86(1) & $-$17.78(1) \\
   80 & $-$18.98(1) & $-$17.90(1) & $-$18.08(1) & $-$17.72(1) & $-$17.61(1) \\
  100 & $-$18.56(1) & $-$17.79(1) & $-$17.90(1) & $-$17.66(1) & $-$17.60(1) \\
  120 & $-$18.34(1) & $-$17.76(1) & $-$17.84(1) & $-$17.68(1) & $-$17.63(1) \\
  160 & $-$18.09(1) & $-$17.73(1) & $-$17.76(1) & $-$17.68(1) & $-$17.65(1) \\
%----------------------------------------------------------------------
\end{tabular}
\end{ruledtabular}
\end{table}

\begin{table}
\caption{\label{Table:C}
Heat capacity (in unit of $k_B$) calculated using the double centroid
virial estimator based on the finite difference scheme in \Sec{fd}.
Other details are the same as in \Table{E}.
}
\begin{ruledtabular}
\begin{tabular}{cccccc}
%----------------------------------------------------------------------
P & PA & TIA & SA($\alpha=0$) & SA ($\alpha$=1/2) & SA($\alpha$=1) \\
%----------------------------------------------------------------------
\hline
   20 &  80.6(4) &  59.3(5) &  65.5(5) &  58.8(5) &  64.2(5) \\
   40 &  55.5(4) &  44.5(5) &  47.4(4) &  42.5(5) &  44.4(5) \\
   60 &  47.7(4) &  38.9(4) &  40.9(4) &  36.9(5) &  36.4(5) \\
   80 &  42.5(4) &  37.7(4) &  38.5(4) &  35.8(4) &  35.1(5) \\
  100 &  40.7(4) &  35.8(4) &  37.3(4) &  35.2(4) &  34.5(4) \\
  120 &  39.3(4) &  35.7(4) &  35.6(4) &  34.9(4) &  34.8(4) \\
  160 &  37.6(4) &  34.6(4) &  35.5(4) &  34.3(4) &  33.6(4) \\
%----------------------------------------------------------------------
\end{tabular}
\end{ruledtabular}
\end{table}

\begin{figure}
\includegraphics[width=7.5cm,clip]{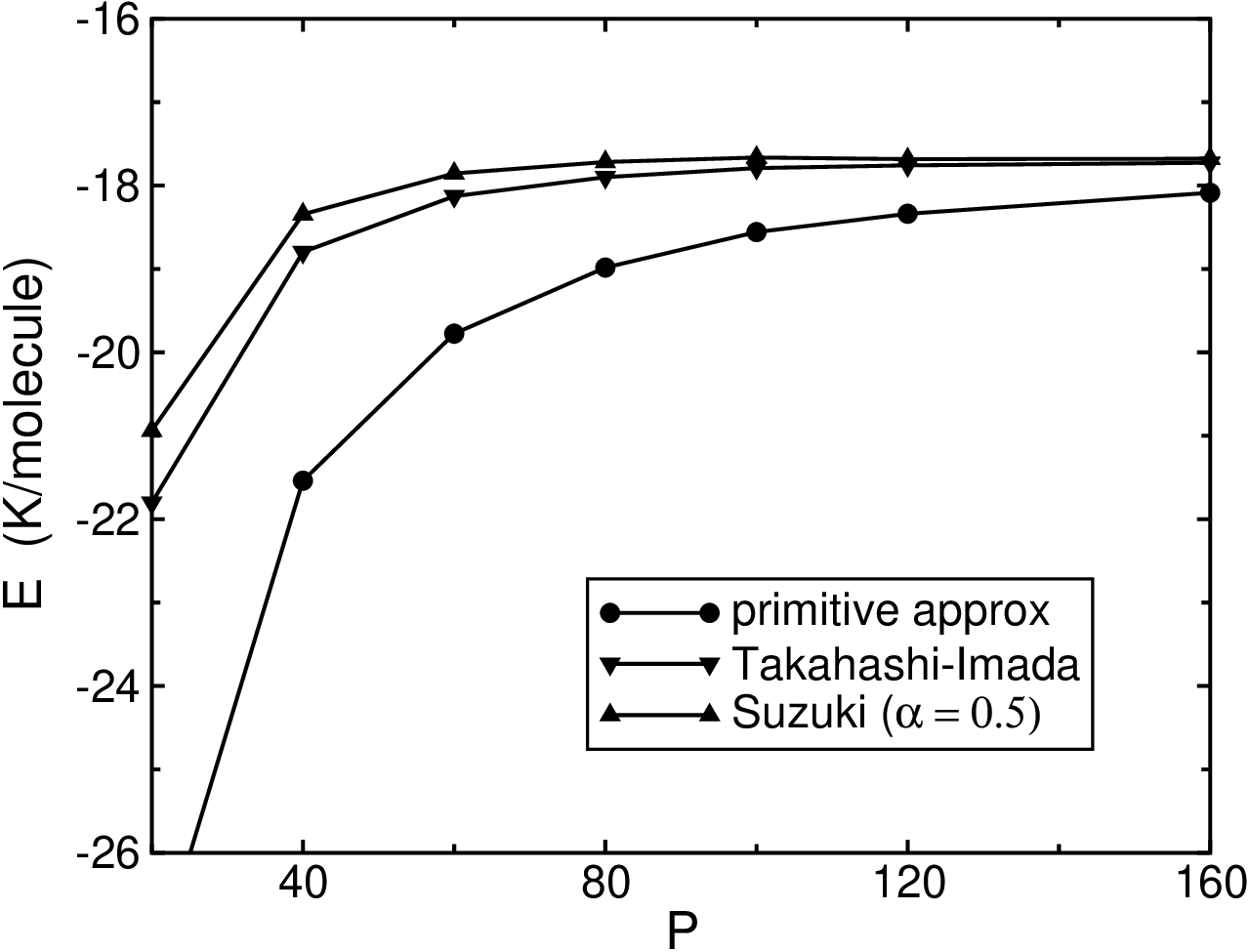}
\includegraphics[width=7.5cm,clip]{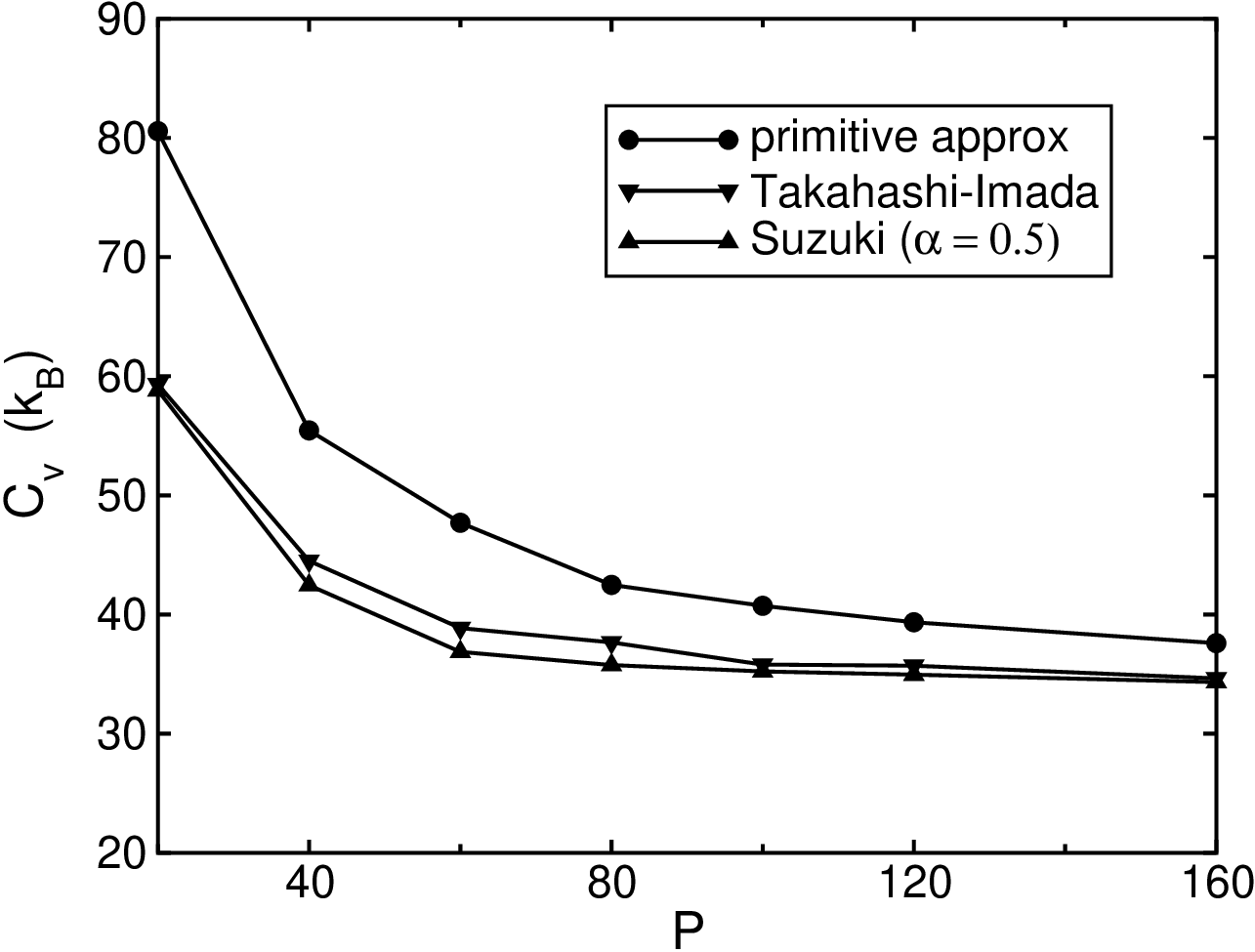}
\caption{\label{Fig:sys_err}
Systematic convergence of (a) energy (K/molecule) and (b) heat capacity
(in unit of $k_B$) of the \hydrogen cluster at 6 K as a function of
the Trotter number $P$.
}
\end{figure}

Tables \ref{Table:E} and \ref{Table:C}
list the energy and heat capacity obtained using
the primitive, Takahashi-Imada, and Suzuki approximations.
\Figure{sys_err} illustrates the systematic convergence of those values
to the $P \rightarrow \infty$ limit.
The centroid-reference virial estimator was used throughout
based on the finite difference scheme presented in \Sec{fd}.
The relative statistical error was estimated to be
on the order of 0.1 and 1 \% for energy and heat capacity, respectively,
by using a blocking procedure with 2000 blocks each of 2000 cycles.
The present calculation needed
only up to first-order potential derivatives as mentioned in the Introduction.
Acceptable values of the stepsize $\delta\beta$
ranged broadly from $10^{-3}\beta$ to $10^{-6}\beta$, where
the smallest value was determined by round-off errors in the heat capacity.
In this paper we set $\delta\beta$
to $10^{-4}\beta$, which practically gave the same result
as when the analytical virial estimator was used.

\Figure{sys_err} shows that the fourth-order approximations
improve remarkably upon the
primitive approximation for both the energy and heat capacity. For
example, to achieve a systematic error in the energy
less than 0.25 K/molecule,\cite{CGC98,DF99,PSDF03_H2}
the primitive approximation requires $P > 200$
while the Suzuki approximation having $\alpha \gtrsim 0.5$ attains
the same accuracy with $P = 60$, thus reducing the necessary value of $P$
by a factor of $\sim$3. This acceleration of systematic convergence
is similar to that observed by Brualla \etal\cite{BSBC04}
in the study of liquid $^4$He at 5.1 K using the Takahashi-Imada approximation.
Regarding the converged values of internal energy, the present result
($E = -17.68 \pm 0.01$ K/molecule)
obtained using the Suzuki approximation with $P = 160$ and $\alpha = 0.5$
is in excellent agreement with the most accurate estimate
($E = -17.69 \pm 0.01$ K/molecule)
obtained by Predescu \etal\cite{PSDF03_H2} using the Wiener-Fourier reweighted
path integral method.
Comparing different fourth-order propagators, we see that
the Suzuki propagator with $\alpha = 0.5$ and 1.0
converges somewhat faster than that with $\alpha = 0.0$ or the Takahashi-Imada
approximation.

\begin{figure}
\includegraphics[width=7.5cm,clip]{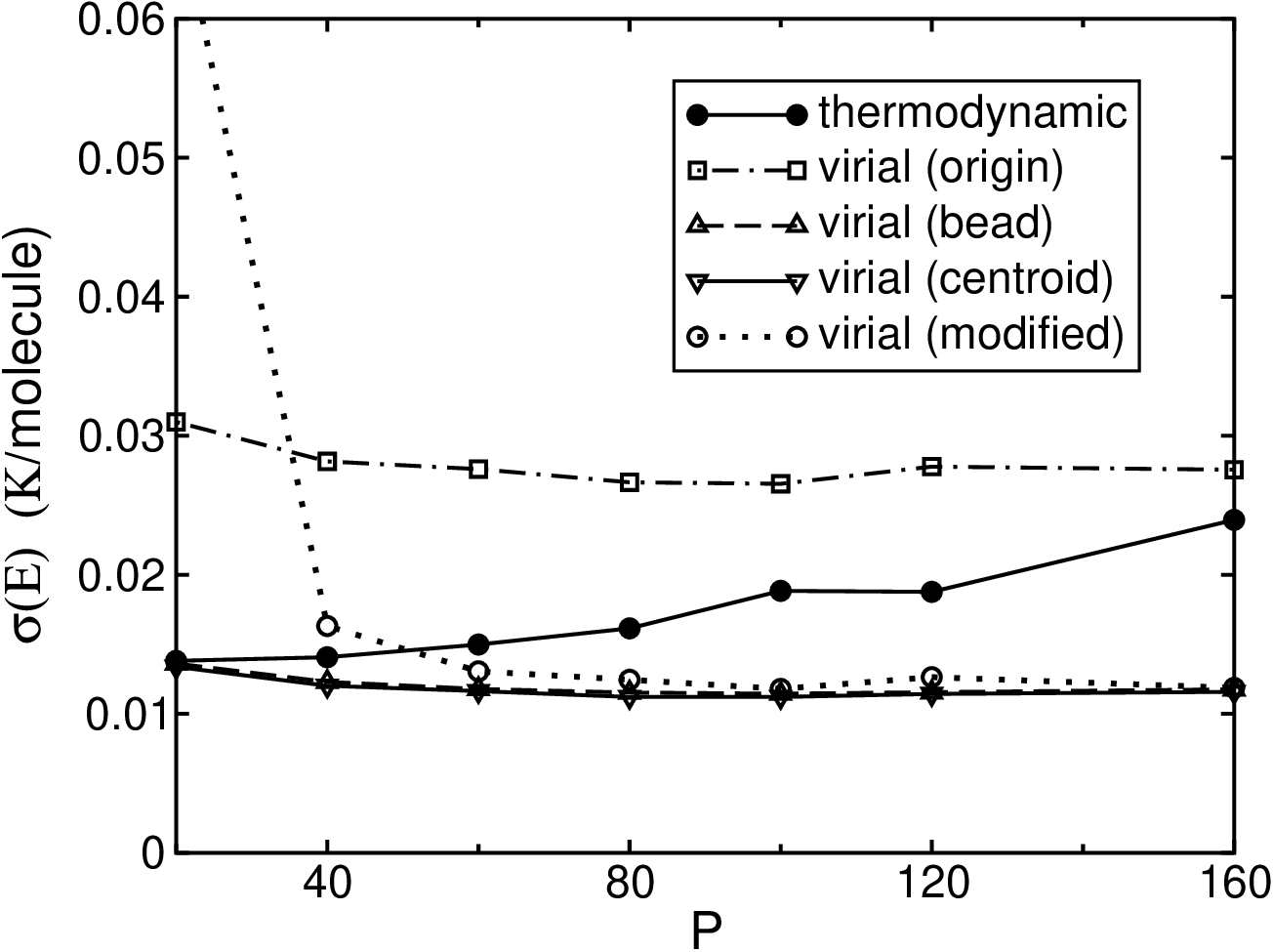}
\includegraphics[width=7.5cm,clip]{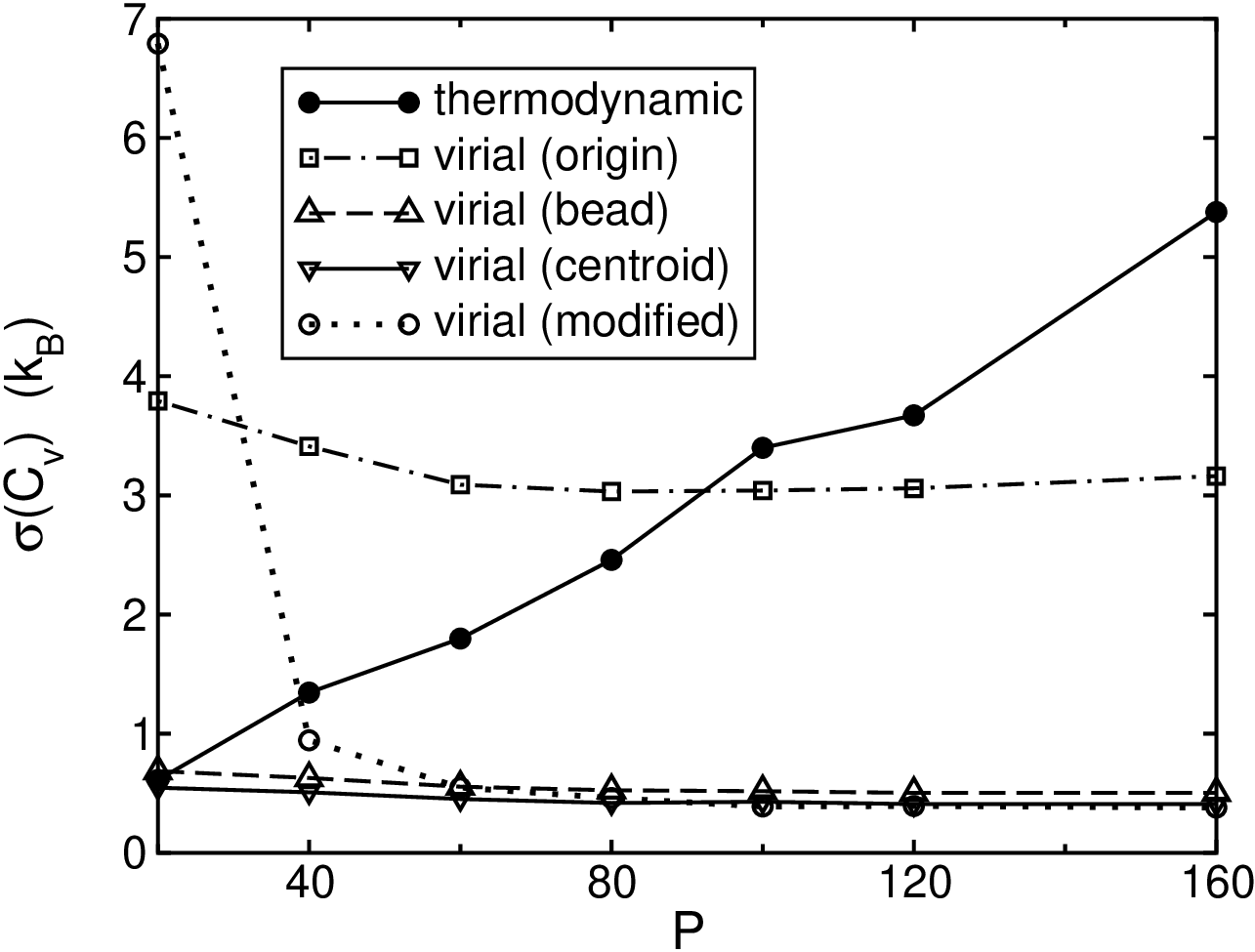}
\caption{\label{Fig:stat_err}
Statistical error in (a) energy (K/molecule) and (b) heat capacity
(in unit of $k_B$) of the \hydrogen cluster at 6 K as a function
of the Trotter number $P$.
Five different estimators are compared: the thermodynamic estimator,
the virial estimator with different choices of the reference point,
and a modified virial estimator in \Eq{mod}.
Errors in the heat capacity were estimated
using the prescription given in \Ref{PSDF03_Ne}.
}
\end{figure}

\Figure{stat_err} compares statistical errors in the energy and heat capacity
obtained with different estimators.
The discretization was performed using the Suzuki approximation with
$\alpha$ = 0.5. This figure shows that
the thermodynamic estimator has a growing variance with $P$
while the virial estimators have a nearly constant variance.
We should note, however, that the variance of the virial estimator
strongly depends on the choice of the reference point $\xvec^*$ in \Eq{scaling}.
That is, the origin-reference virial estimator
exhibits a significantly larger variance than the bead- or
centroid-reference estimators, indicating that it is more advantageous
to choose $\xvec^*$ in \Eq{scaling} as $\xvec_P$ or $\xcent$
than the coordinate origin.

\Figure{stat_err} also plots the statistical error
of the \textit{modified} centroid virial estimator that may be used
in path integral molecular dynamics.\cite{PR84,TBMK93}
In the latter method the fourth-order composite propagators become
expensive if $\rho^{(4)}$ in \Eq{rho_FOA} is used directly
as a sampling function, because the ``forces'' exerted on the beads
require second-order potential derivatives. This problem can be
avoided, for example, by excluding the force square terms
in \Eq{commutator} from the sampling function.\cite{JJV01}
The resulting modified expression for the energy is
\begin{equation}
\label{Eq:mod}
  E ( \beta )
  \simeq
  - \frac{ \langle \Delta\rho \varepsilon_V \rangle_\mathrm{mod} }
         { \langle \Delta\rho \rangle_\mathrm{mod} }
\end{equation}
with
\begin{equation}
  \Delta \rho = \exp \left\{ - \sum^P_{s = 1} w_s d_s ( \beta / P )^3 C ( \xvec_s
  ) \right\},
\end{equation}
where various symbols are the same as in \Sec{FOA}. $\langle \cdots
\rangle_{\mathrm{mod}}$ in \Eq{mod} denotes an ensemble average over
the following sampling function:
\begin{equation}
\label{Eq:rho_mod}
  \rho_{\mathrm{mod}} ( \xvec_1, \ldots, \xvec_P )
  =
  \exp \left\{ - \frac{m P}{2 \hbar^2 \beta}
  \sum^P_{s = 1} ( \xvec_s - \xvec_{ s-1 } )^2
  - \frac{\beta}{P} \sum^P_{s = 1} w_s V ( \xvec_s ) \right\} .
\end{equation}
This method gives the true expected value of energy as $P$ is increased, but
the statistical error becomes larger than the original scheme in
\Sec{method}.\cite{JJV01}
\Figure{stat_err} shows that the variance obtained with this method
is quite large for small values of $P$ but is reduced to a manageable
size if $P$ is increased to $>$40.
Thus, excluding force square terms from the
sampling function seems a viable option
if molecular dynamics methods are used as a statistical sampler.

\begin{figure}
\includegraphics[width=7.5cm,clip]{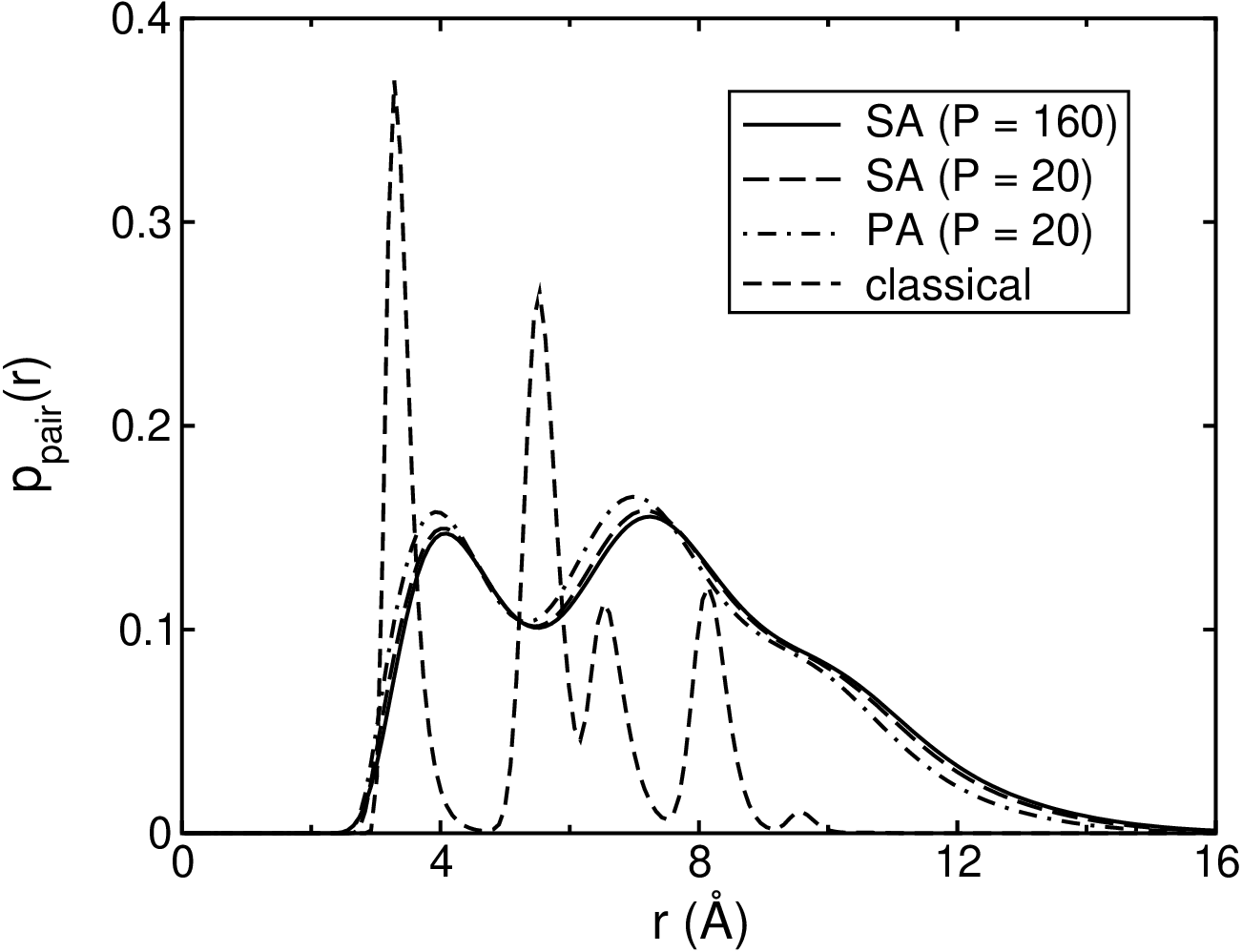}
\includegraphics[width=7.5cm,clip]{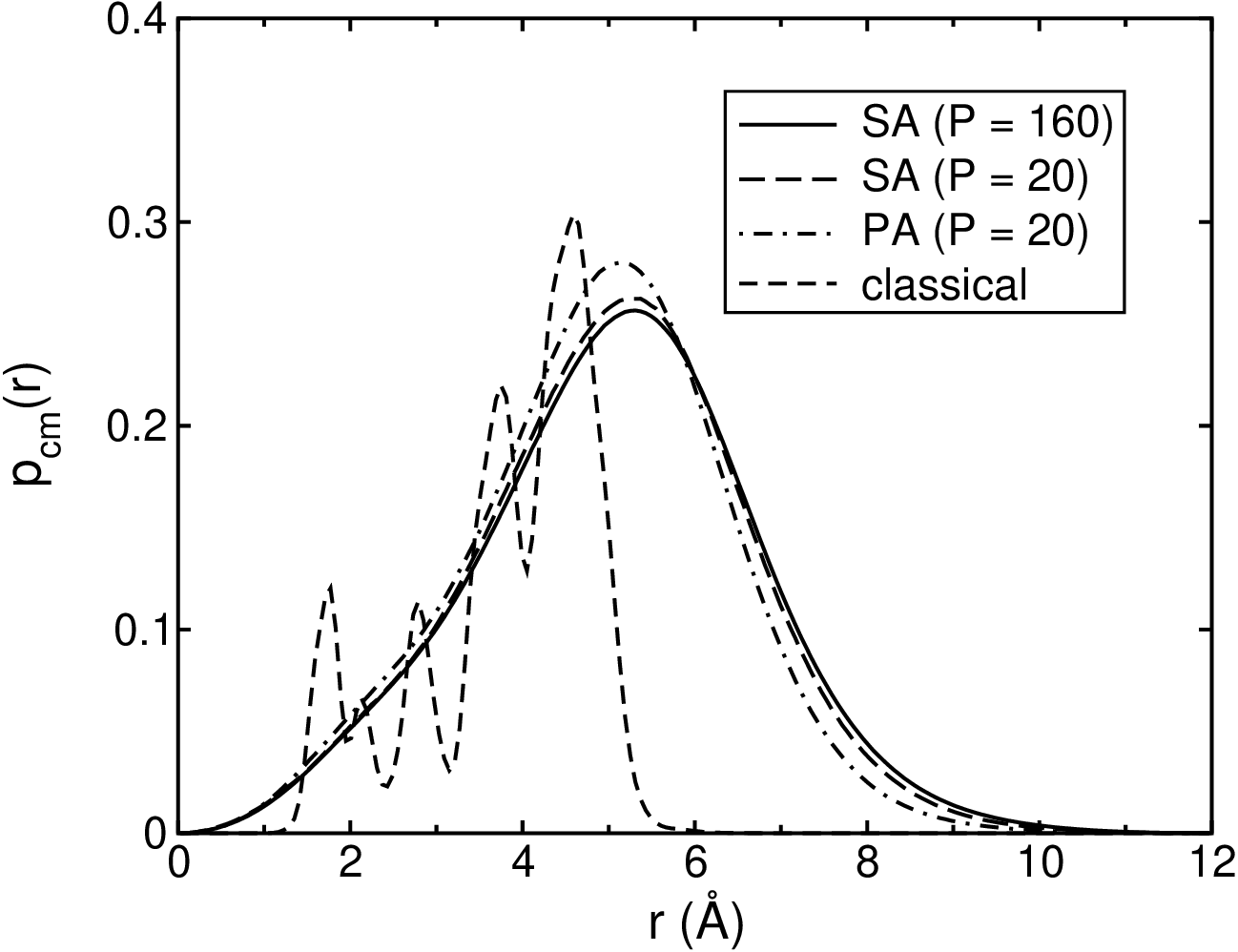}
\caption{\label{Fig:rdf_hydrogen}
(a) Pair radial distribution function and (b) distance distribution function
from the center of mass of the \hydrogen cluster at 6 K.
SA and PA denote the quantum results obtained with the Suzuki and primitive
approximations, respectively. The classical result is plotted
in short dashed line (with its height scaled by a factor of 1/2
to fit in the panel).
}
\end{figure}

\Figure{rdf_hydrogen} illustrates the classical and quantum results of
the pair radial distribution function and
the distance distribution function from the cluster center of mass defined by
\begin{equation}
  \label{Eq:p_pair}
  p_\mathrm{pair}( r )
  \propto
  \left\langle
     \sum_{ i < j }^{ N }
     \sum_{ s }
     \delta ( r - r_{ij}^{(s)} )
  \right\rangle
  ,
\end{equation}
and
\begin{equation}
  \label{Eq:p_cm}
  p_\mathrm{cm}( r )
  \propto
  \left\langle
     \sum_{ i = 1 }^{ N }
     \sum_{ s }
     \delta ( r - \Delta r_i^{(s)} )
  \right\rangle
  ,
\end{equation}
respectively, where $r_{ij} = | \rvec_i - \rvec_j |$
and $\Delta r_i = | \rvec_i - \Rvec |$.
The sum over time slices is performed for all (only even)
values of $s$ when the primitive (Suzuki) approximation is used.\cite{JJV01}
Here we do not consider the Takahashi-Imada approximation
because it requires a nontrivial modification
to the estimator.\cite{KM02}
We see from \Fig{rdf_hydrogen} that the classical cluster 
has a rigid, solidlike structure at this temperature,\cite{replica_exchange} while
the quantum cluster has a liquidlike structure due to large
zero-point energies and tunneling effects.
This figure also shows that both the primitive and Suzuki
approximations with $P = 20$ already give a good approximation
to the practically exact result obtained with $P = 160$,
indicating that structural properties converge much faster
than the energy and heat capacity as a function of $P$.

%--------------------------------------------------

\subsection{Ne$_{13}$ cluster at 4--14 K}

\neon is one of the smallest clusters that exhibit
solid-liquid-like (or melting) transition,
and it has been studied extensively using
a variety of theoretical methods.\cite{NFD00,PSDF03_Ne,C95,CDW01,B02,FM04,PFM05}
The classical melting point is located at around 10 K and it is
lowered by about 10 \% due to prominent quantum effects.
The heat capacity is a useful quantity for characterizing
such a cluster phase transition.
Neirotti \etal\cite{NFD00} calculated the heat capacity of \neon
using a double virial estimator (in analytical form) designed for
the Fourier path integral, but unfortunately
their results exhibited a large statistical error (about 10 $k_B$)
in the low temperature region.
Predescu \etal\cite{PSDF03_Ne} calculated the same quantity
using their viriallike estimator (in finite-difference form)
in the framework of random series
path integral, and as mentioned in the Introduction
they obtained highly converged results
with statistical errors less than 1 $k_B$.
Because the two calculations used the same number of Monte Carlo samples,
this reduction in statistical error corresponds roughly to
100 times acceleration in convergence rate.
Then a natural question that arises is what is
the dominant factor that reduced the statistical error.
We find, however, that this question is rather difficult to answer
because there are quite a few technical differences in their calculations.

As such, to get some insights into the above question,
we have re-calculated the heat capacity of \neon
using the discretized path integral.
The computational details are basically the same as in the preceding section,
and the relevant parameters were set as closely as possible
to those in \Refs{NFD00}{PSDF03_Ne}.
Specifically, the Lennard-Jones parameters were set to
$\epsilon_\mathrm{LJ} = 35.6$ K and $\sigma_\mathrm{LJ} = 2.749$ \AA, and
the mass of Ne was 20.0 amu. The confining radius $R_c$ in \Eq{V_c}
was chosen as $2 \sigma_\mathrm{LJ}$.
The number of Monte Carlo cycles was $4\times10^6$.
We also performed the replica-exchange (or parallel tempering)
Monte Carlo\cite{MP92,HN96,H97} to avoid nonergodicity problem
at low temperature.
The number of replicas was set to 21, and the replica temperatures
were distributed over the interval [4,14] K with even spacing.
The exchange move was attempted every 10 Monte Carlo cycles.
This setting ensured
the acceptance ratio of exchange moves to be $>$10 \%.

\begin{figure}
\includegraphics[width=7.5cm,clip]{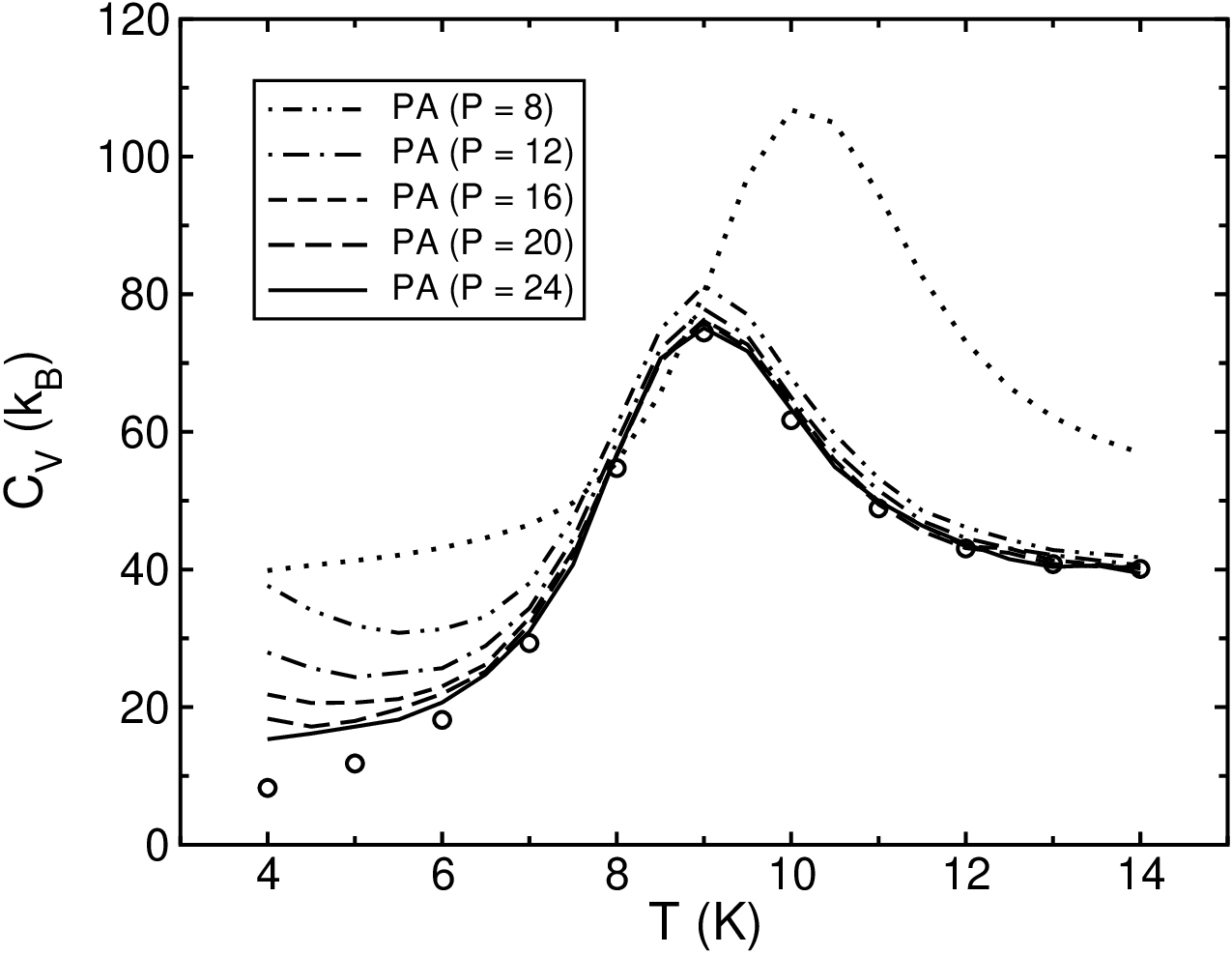}
\includegraphics[width=7.5cm,clip]{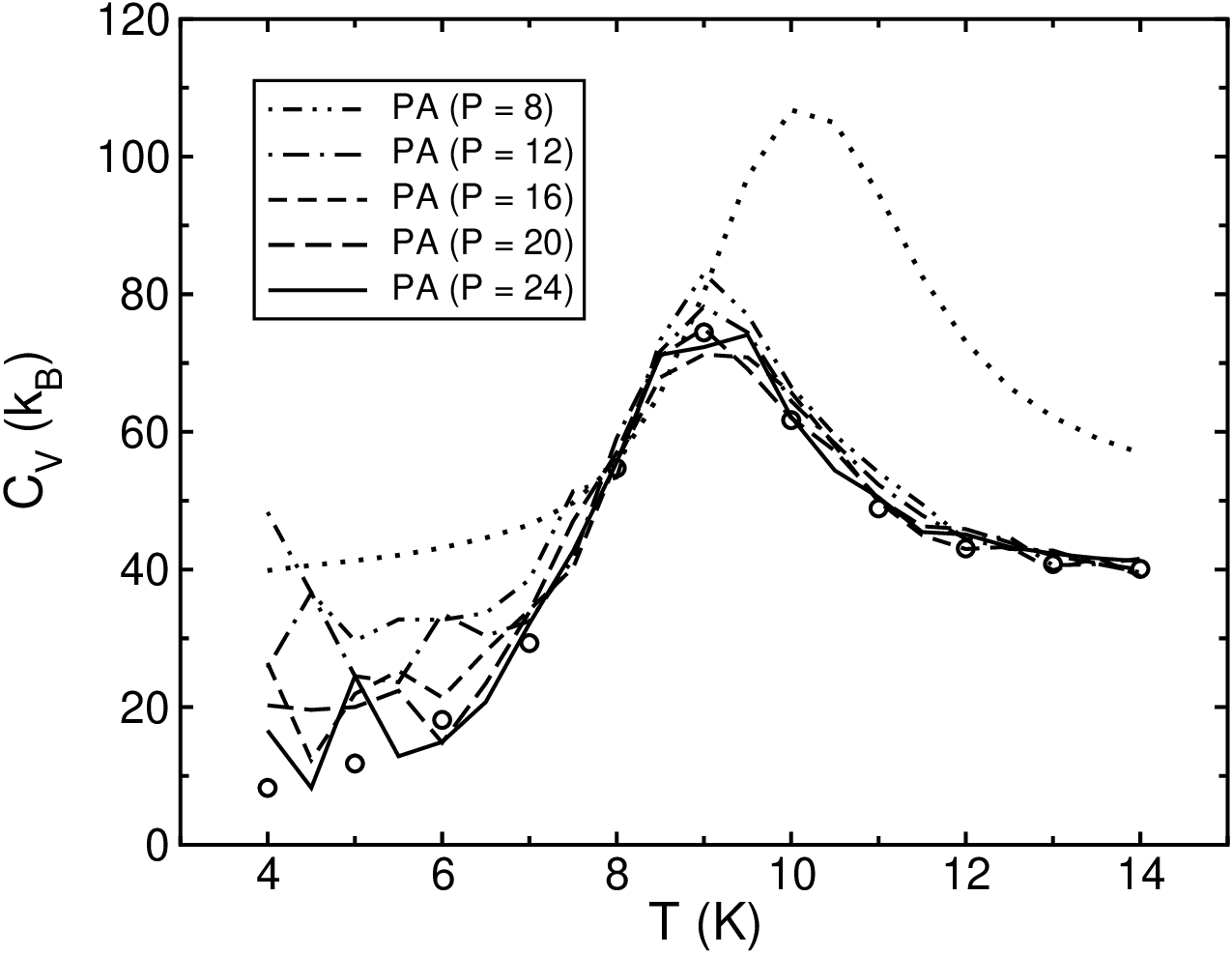}
\includegraphics[width=7.5cm,clip]{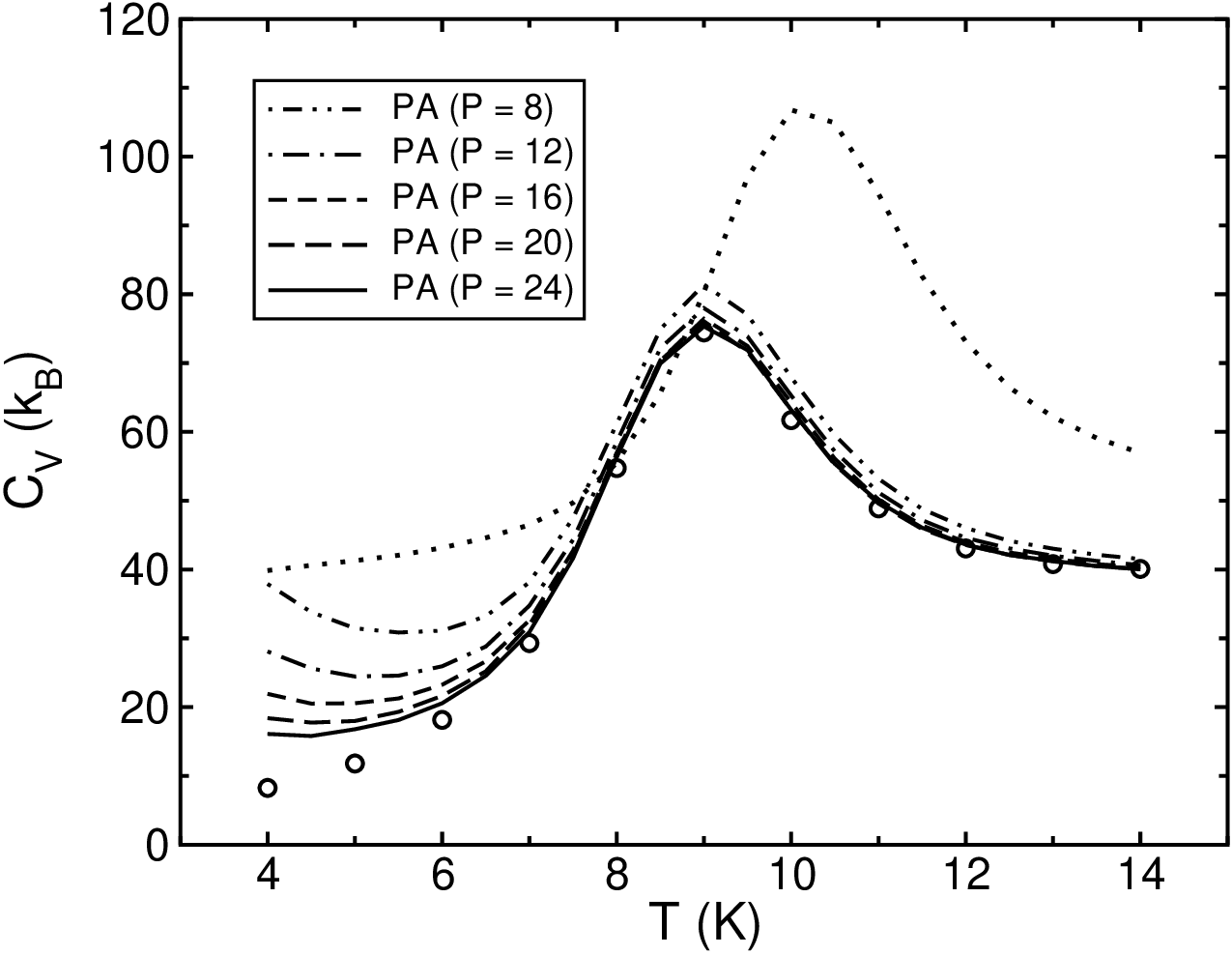}
\includegraphics[width=7.5cm,clip]{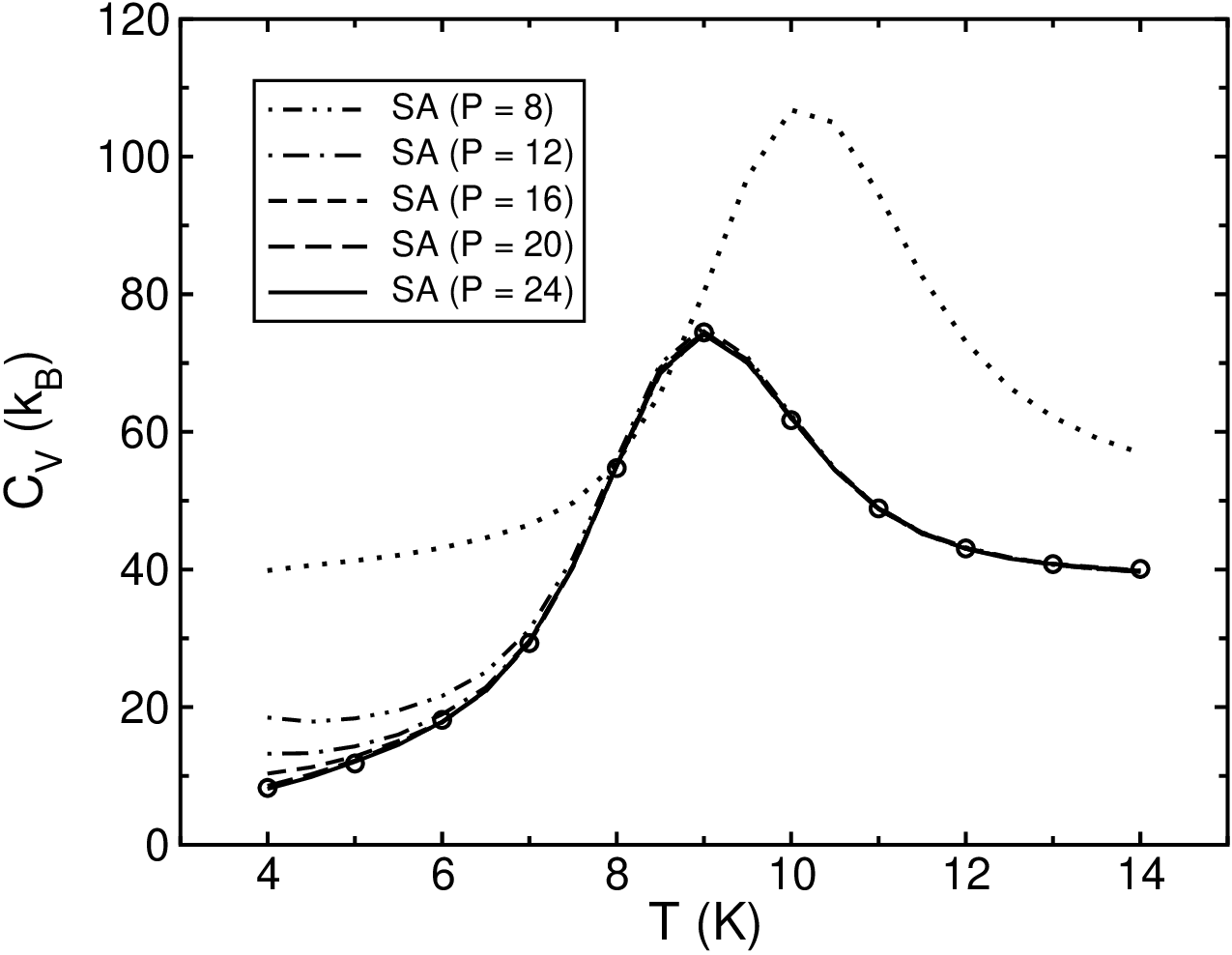}
\caption{\label{Fig:C_neon}
Heat capacity of the \neon cluster as a function of temperature.
SA and PA denote the quantum results obtained with the Suzuki and primitive
approximations, respectively. Different estimators are used in each panel:
(a) double thermodynamic estimator in \Eq{double_TD};
(b) origin-reference double virial estimator;
(c) bead-reference double virial estimator;
(d) centroid-reference double virial estimator.
The statistical errors in panels (a), (c), and (d) are comparable
to the width of the line while that in panel (b) is about
5 $k_B$ in the low temperature region.
The highly accurate results obtained by Predescu \etal~(\Ref{PSDF03_Ne}) are
plotted by circles. The classical result is plotted in dotted line.
}
\end{figure}

\Figure{C_neon} plots the heat capacity thus obtained as a function of temperature.
Four combinations of the approximate propagator and the heat capacity estimator
are examined, namely:
\begin{enumerate}[(a)]
\item primitive approximation + double thermodynamic estimator;
\item primitive approximation + origin-reference double virial estimator;
\item primitive approximation + bead-reference double virial estimator;
\item Suzuki approximation ($\alpha=0.5$)
      + centroid-reference double virial estimator.
\end{enumerate}
In cases (b), (c), and (d) the virial estimator was evaluated
using the finite difference scheme in \Sec{fd}, while in case (a)
the heat capacity was calculated using the double thermodynamic estimator:
\begin{equation}
   \label{Eq:double_TD}
   C_V( \beta )
   =
   k_B \beta^2
   \left\{
      \langle \epsT^2 \rangle
      -
      \langle \epsT \rangle^2
      -
      \langle \frac{ \del \epsT }{ \del \beta } \rangle
   \right\}
   ,
\end{equation}
where $\epsT$ is given by \Eq{eps_T}. In all cases the number of
time slices was set to $P$ = 8--24. Also shown in \Fig{C_neon}
is the highly accurate results obtained by Predescu
\etal\cite{PSDF03_Ne} (circles) and the classical heat capacity
(dotted line). This figure reveals that
the origin-reference virial estimator has much larger statistical
errors than the other three cases. Also interesting is the fact that
the variance of the double thermodynamic estimator is rather small and
close to the bead- and centroid-reference double virial estimators.
This tendency is qualitatively similar to that observed for
the heat capacity of the hydrogen cluster in \Fig{stat_err},
where the origin-reference virial estimator
has the largest statistical error in the small $P$ region.
Regarding the systematic convergence to the $P \rightarrow \infty$ limit,
\Fig{C_neon} (d) shows that the Suzuki propagator again provides
a noticeable improvement over the primitive approximation, and that $P = 20$ is
sufficient to reach systematic convergence within 1 $k_B$.

What is more important about the question discussed above is that
cases (b) and (c) correspond qualitatively to the calculation
by Neirotti \etal\cite{NFD00} and
Predescu \etal,\cite{PSDF03_Ne} respectively.
More precisely, the double virial estimator of
Neirotti \etal~may be regarded as origin-reference because their estimator
vanishes when the interaction potential is set to 0 (and thus the contribution
of the center of mass was treated separately),
while the finite-difference
estimator of Predescu \etal~may be viewed as bead-reference
because the relevant Brownian bridge is defined in terms of
a set of ``physical coordinates'' (equivalent to a single bead).
Thus, we think that the dominant factor that made a large difference
in their calculations is the choice of the reference point in the virial estimator,
rather than whether the estimator
was evaluated analytically\cite{NFD00} or numerically
via finite difference\cite{PSDF03_Ne}
if we consider the fact that cases (b), (c), and (d) above
were treated using the finite difference method in \Sec{fd}.

\begin{figure}
\includegraphics[width=7.5cm,clip]{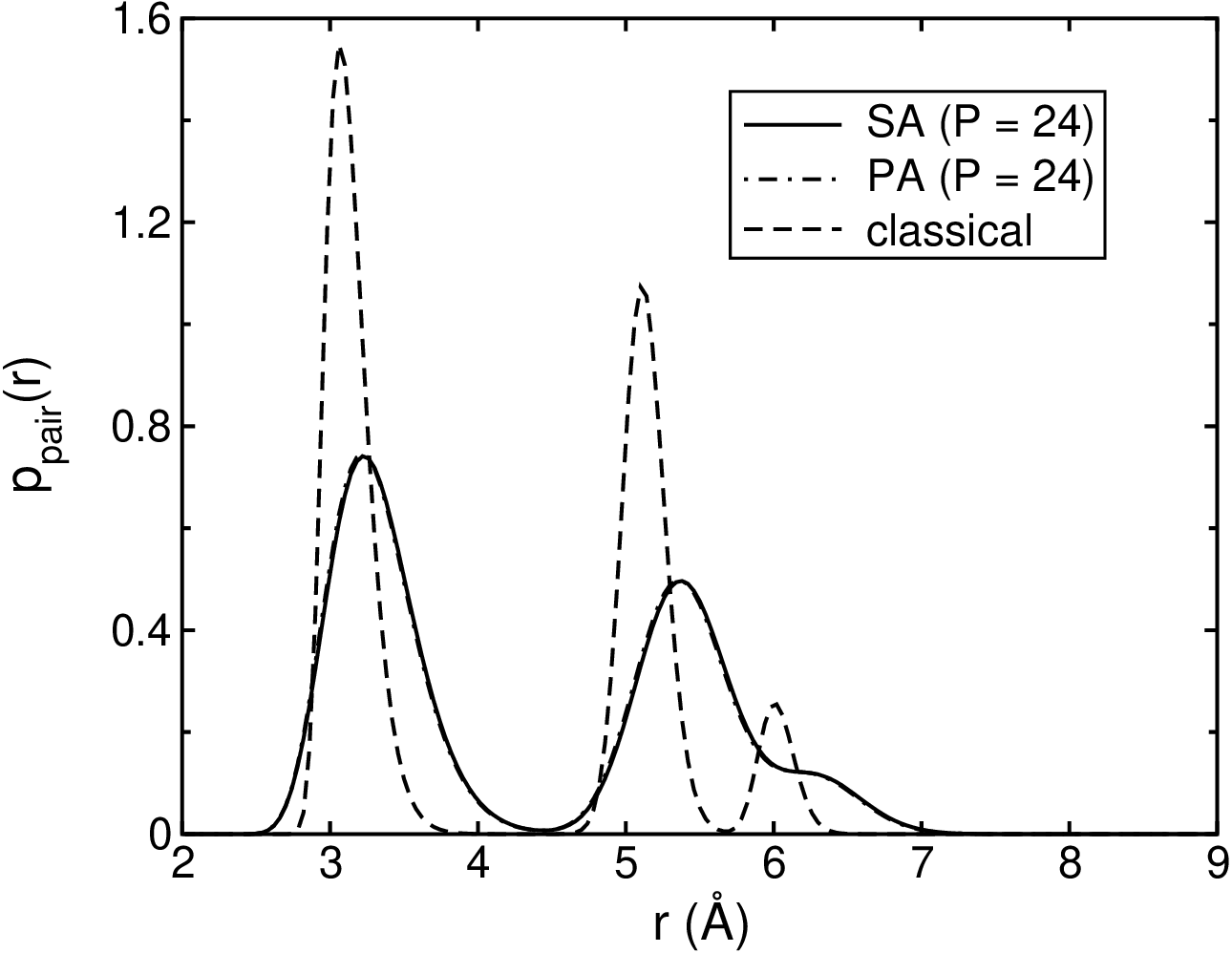}
\includegraphics[width=7.5cm,clip]{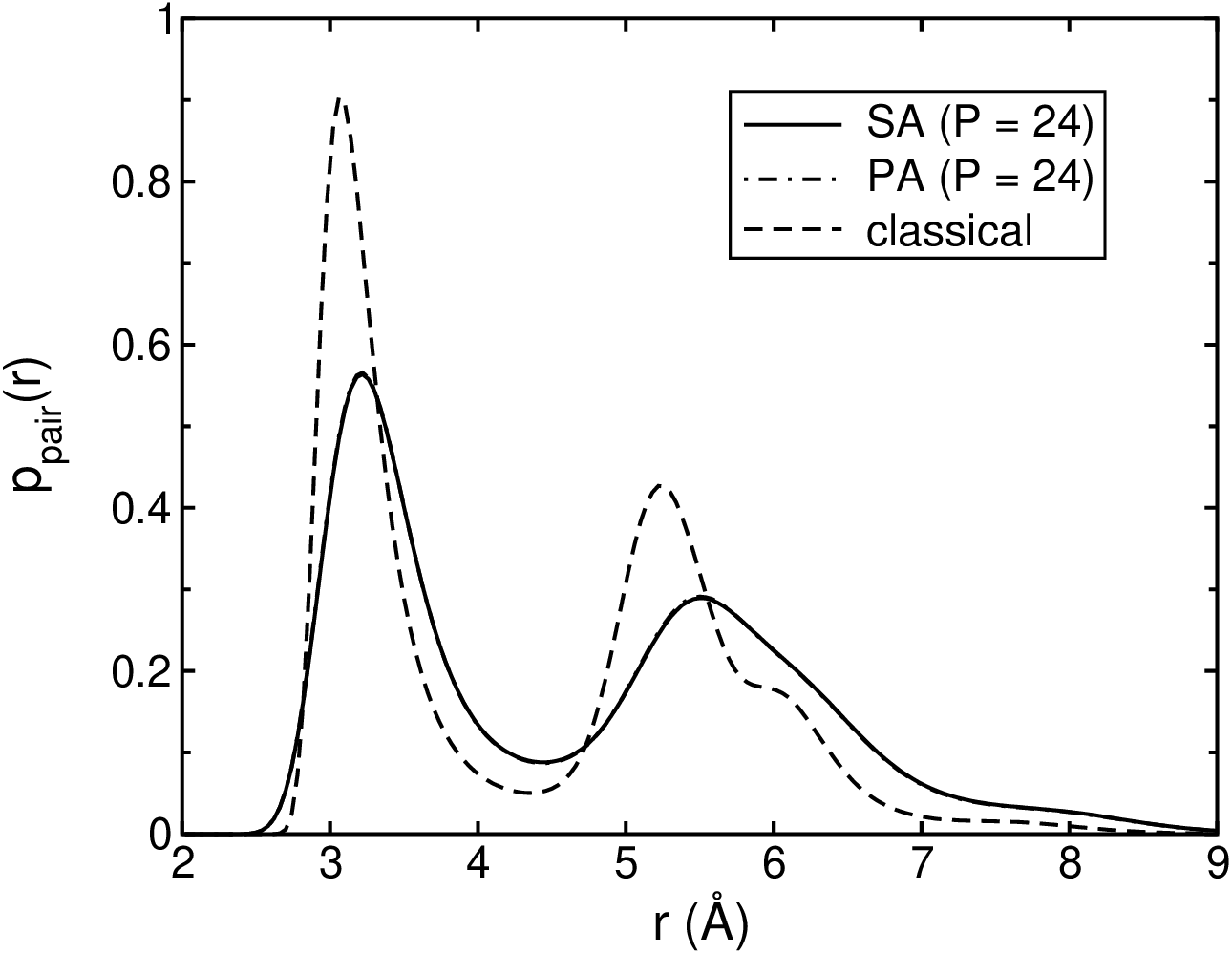}
\caption{\label{Fig:rdf_neon}
Pair radial distribution functions of the \neon cluster at (a) 4 K and (b) 10 K.
SA and PA denote the quantum results obtained with the Suzuki and primitive
approximations, respectively. The classical result
is plotted in dashed line.
}
\end{figure}

Finally, \Fig{rdf_neon} illustrates the pair
distribution functions at 4 and 10 K, which
are very similar to those presented in \Refs{B02}{FM04}.
The cluster takes a solidlike structure at 4 K while
it starts to form a liquidlike structure at 10 K
(slightly above the melting temperature).
Comparing the classical and quantum results,
we see that the positions of the classical peaks
are shifted outward and their widths
broadened when quantum effects are made operative.
The degree of broadening is much smaller than that observed for the hydrogen
cluster in \Fig{rdf_hydrogen} due to quasiclassical nature of Ne$_{13}$.
\Figure{rdf_neon} also shows that the primitive and Suzuki approximations
with $P = 24$ give almost indistinguishable results, verifying
the fast convergence of structural properties with respect to $P$.

%----------------------------------------------------------------------

\section{Conclusions\label{Sec:conclusion}}

In this paper we have presented a coordinate scaling procedure for obtaining
the conventional virial estimator and discussed
its efficient evaluation using finite difference with respect to
temperature. This procedure allowed us to apply
the fourth-order propagators to quantum clusters
using only the first-order potential derivatives. From the results of the
application, we find that setting the reference point in the virial estimator
to the coordinate origin (the path centroid) gives the largest (smallest)
statistical errors. This result is also in qualitative agreement
with previous studies on Ar clusters and liquid water.\cite{GF02a,GF02b,SS05}
Thus, despite its extensive use in the literature,
it is \textit{not} recommended to use the origin-reference
virial estimator in quantum clusters and condensed phase systems
because of the large variance as well as unnecessary
complication due to unbounded degrees of freedom.

We end this paper by mentioning some possible application
of the present method.
One example is a short-time approximation
to the quantum correlation function $C_{A B} ( t )$, e.g.,
\begin{equation}
  C_{A B} ( t ) = C_{A B} ( 0 ) + \frac{1}{2} \ddot{C}_{A B} ( 0 ) t^2 +
  \cdots \simeq C_{A B} ( 0 ) \exp \left[ \frac{1}{2} \frac{\ddot{C}_{A B} ( 0
  )}{C_{A B} ( 0 )} t^2 \right]
  .
\end{equation}
Taking the real-time derivatives of $C_{AB}(t)$ at $t = 0$
along the imaginary-time axis results in a path integral calculation
similar to that of heat capacity,\cite{YM04}
which implies a similar reduction in statistical errors via coordinate
scaling. Such an idea is particularly relevant, e.g.,
to an approximate calculation of chemical reaction rates\cite{YM04,ZYM04,YM05,P04b}
or vibrational relaxation rates.\cite{RR01}
Another possible application is the
path integral ground state (or variational
path integral) methods,\cite{DMC95,SSM00,CRB05}
where several different estimators arise in natural analogy
to finite temperature path integral.

%----------------------------------------------------------------------

%
% References
%

\newcommand{\JCP}[1]{J. Chem. Phys. \textbf{#1}}

\newcommand{\JPCB}[1]{J. Phys. Chem. B \textbf{#1}}

\newcommand{\PR}[2]{Phys. Rev. #1 \textbf{#2}}

\bibliography{basename of .bib file}

\begin{thebibliography}{99}
\bibitem{int_by_parts}
For $\xvec^* = \xvec_P$ the $P$-th term in the square bracket in \Eq{int_by_parts}
vanishes, so that integration by parts of the remaining $P-1$ terms gives $g =
1$. The result for $\xvec^* = \xcent$ can be obtained via cyclic averaging of a
similar equation with $\xvec^* = \xvec_t$, $t = 1, \ldots, P$.
\bibitem{Euler_theorem}
This can be shown by differentiating the relation
$K( \{ \xvec^* + \lambda ( \xvec_s - \xvec^* ) \} ) = \lambda^2 K(\{ \xvec_s \})$
with respect to $\lambda$ and setting $\lambda$ to 1, where $K(\{ \xvec_s \})$
is the kinetic action in \Eq{rho_PA}.
\bibitem{smooth_potential}
In this paper we consider only smooth potentials that have continuous first- and
second-order derivatives.
\bibitem{replica_exchange}
In the classical calculation we used the replica-exchange
Monte Carlo to avoid possible nonergodicity problem in the solid-like
phase at low temperature. The replica temperatures were
distributed over [6,30] K with even spacing of 1 K, which
ensured $>$40 \% acceptance ratio for the exchange move.
\end{thebibliography}

\end{document}